\documentclass[11pt]{revtex4-1}

\usepackage{graphicx}

\begin{document}

\title{Do recent observations favor a cosmological event horizon: A thermodynamical prescription?}

\author{Subhajit Saha\footnote {subhajit1729@gmail.com}}
\author{Subenoy Chakraborty\footnote {schakraborty.math@gmail.com}}
\affiliation{Department of Mathematics, Jadavpur University, Kolkata-700032, West Bengal, India.}

\begin{abstract}
Recent observational evidences claim an accelerating expansion of the universe at present epoch. It is commonly incorporated in standard cosmology by the introduction of an exotic matter (that violates the strong energy condition) known as dark energy (DE). As event horizon exists for accelerating universe so there has been a lot of work on universal thermodynamics ({\it i.e.,} thermodynamics of the universe bounded by apparent or event horizon). Recently, thermodynamical equilibrium has been examined for both the horizons. In the present work we show that universal thermodynamics with event horizon is favored by DE from the point of view of equilibrium thermodynamical prescription.

\end{abstract}

\maketitle
\section{INTRODUCTION}

Almost at the fag end of the last century there was a great blow to the standard cosmology due to the observational prediction [1-3] that the universe is passing through an accelerating phase. As a result the standard cosmology was tuned either physically by incorporating dominant matter component having large negative pressure (known as dark energy (DE)) or geometrically, by introducing modified gravity theories.

In the recent past, a lot of work has been going on universal thermodynamics, mostly with apparent horizon as the boundary. The pioneer work of Wang {\it et al.} [4] has made a comparative study of the two horizons (apparent and event) by examining the validity of the thermodynamical laws for DE fluids. Based on their results, they have concluded that universe bounded by the apparent horizon is a Bekenstein system while cosmological event horizon is unphysical from thermodynamical point of view. However, due to the relevance of the cosmological event horizon in the present DE dominated universe, we have shown [5] the validity of the generalized second law of thermodynamics (GSLT) on the event horizon (in any gravity theory) assuming the validity of the first law with some reasonable restrictions. Subsequently, it was possible to show [6] the validity of the first law of thermodynamics on the event horizon for some simple DE fluids with a modified Hawking temperature on the event horizon. Also very recently, thermodynamical equilibrium has been studied [7,8] for universe bounded by any of the two horizons and for various DE models. Motivated by one of the latest work of Radicella and Pavon [9] on the thermodynamic motivation for DE, in the present work, we make an attempt to show that universe bounded by event horizon is favoured by DE fluid from thermodynamical point of view {\it i.e.,} validity of GSLT as well as thermodynamical equilibrium.

According to thermodynamics, the equilibrium configuration of an isolated macroscopic physical system should be the maximum entropy state (consistent with the constraints imposed on the system). Thus if S be the entropy of the system then we must have [7,10]\\\\
i) $\dot{S} \geq 0$ ({\it i.e.,} the entropy function cannot decrease, the second law of thermodynamics)\\
ii)$\ddot{S} < 0$ ({\it i.e.,} the entropy function attains a maximum).\\

In the present context, the universe filled with DE fluid and bounded by the horizon is considered as an isolated macroscopic system for which the above inequalities are generalized as [7]
\begin{equation}
i)~\dot{S}_h+\dot{S}_{fh} \geq 0~~~and~~~ii)~\ddot{S}_h+\ddot{S}_{fh} < 0,
\end{equation}
where $S_h$ and $S_{fh}$ are the entropies of the horizon and that of the fluid within it respectively. 

Now in order to determine the entropy variations we shall use the simplest form of the first law of thermodynamics ({\it i.e.,} Clausius relation),
\begin{equation}
T_h dS_h=\delta Q_h=-dE_h
\end{equation}
and the Gibb's relation [11]
\begin{equation}
T_f dS_{fh}=dE_f+pdV_h,
\end{equation}
where $E_h$ is the energy flow across the horizon, $E_f=\rho V_h$ is the total energy of the fluid inside, $V_h=\frac{4}{3} \pi R^{3}_h$ is the volume of the fluid and ($T_h$, $T_f$) are the temperature of the horizon and that of the fluid respectively.

Here due to the WMAP data [12], we choose flat homogeneous and isotropic FRW model of the universe for which the two horizons are related by the relation
\begin{equation}
R_A=\frac{1}{H}=R_H<R_E,
\end{equation}
where $R_A$ and $R_E$ are the radius of the apparent horizon and that of the event horizon respectively ($R_H$ is the Hubble horizon). As universe bounded by the apparent horizon is a Bekenstein system so we use Bekenstein's entropy-area relation and Hawking temperature as
\begin{equation}
S_A=\frac{\pi R^{2}_A}{G}~~~,~~~T_A=\frac{1}{2\pi R_A}.
\end{equation}
On the other hand, following [6] the entropy and temperature on the event horizon are taken as
\begin{equation}
S_E=\frac{\pi R^{2}_E}{G}~~~,~~~T_E=\frac{H^2R_E}{2\pi}.
\end{equation}
Now due to inequality (4) we have
\begin{equation}
T_A<T_E.
\end{equation} 

In Ref. [8], it has been shown that although GSLT is satisfied for any fluid distribution bounded by the apparent horizon but equilibrium criteria is not satisfied in the quintessence era even for perfect fluid with constant equation of state. However, for universe bounded by the event horizon both GSLT and equilibrium configuration are satisfied with some realistic conditions. In the next section, the thermodynamical prescriptions are studied in details for some specific choices of DE models, for apparent/event horizon and it is examined whether DE favours one of the horizons or not.\\\\\\

\section{THERMODYNAMICAL PRESCRIPTIONS FOR SOME DE MODELS}

In the present section, we shall make a comparative study of the above thermodynamical equilibrium conditions ({\it see} Eq. (1)) for universe bounded by the apparent/event horizon for the following realistic DE fluid models:\\
i)Perfect fluid with constant equation of state: $p=\omega \rho$, $\omega <-\frac{1}{3}$\\
ii)Interacting holographic DE model.\\
iii)Modified Chaplygin Gas: a unified dark matter (DM)-DE model.

\subsection{Perfect fluid with constant equation of state: $p=\omega \rho$, $\omega<-\frac{1}{3}$}

Following Ref. [8] we have

\begin{equation}
\dot{S}_A+\dot{S}_{fA}=\frac{9\pi}{2GH}(1+\omega)^2 =\frac{2\pi}{GH} v_A^{2},
\end{equation}
\begin{equation}
\ddot{S}_A+\ddot{S}_{fA}=\frac{9\pi}{2G}(1+\omega)[(1+\omega)^2+2{\omega}^2]=\frac{4\pi}{3G}v_A[v_A^{2}+2(v_A-\frac{3}{2})^2];
\end{equation}
and
\begin{equation}
\dot{S}_E+\dot{S}_{fE}=\frac{8{\pi}^2 R_E\rho(1+\omega)}{H}(R_E-\frac{1}{H})=\frac{2\pi R_E}{G}v_Av_E,
\end{equation}
\begin{equation}
\ddot{S}_E+\ddot{S}_{fE}=8{\pi}^2\rho(1+\omega)(R_E-\frac{1}{H})[-\lbrace{\frac{R_E}{2}(1+3\omega)+\frac{1}{H}\rbrace}+R_E(1-\frac{v_A}{v_E})]=-\frac{2\pi}{G}v_Av_E[\frac{v_A}{v_E}(1+{v_E})^2-(1+2v_E)];
\end{equation}
where $v_A=\dot{R}_A=\frac{3}{2}(1+\frac{p}{\rho})~~and~~v_E=\dot{R}_E=HR_E-1$ are the velocities of the apparent and the event horizon respectively. From Eqs. (8) and (9), we have for the apparent horizon:\\
$\bullet$ $\dot{S}_A+\dot{S}_{fA}\geq 0$~~for all $\omega$({\it i.e.,} for all $v_A$).\\
$\bullet$ $\ddot{S}_A+\ddot{S}_{fA}>0$~in quintessence era($\omega>-1, {\it i.e.,}~ v_A>0$),\\
$~~~~~~~~~~~~~~~$ $=0$~at phantom crossing($\omega=-1, {\it i.e.,}~ v_A=0$),\\
$~~~~~~~~~~~~~~~$ $<0$~in phantom domain($\omega<-1, {\it i.e.,}~ v_A<0$).\\
Thus the generalized second law of thermodynamics (GSLT) holds both in the quintessence era and in the phantom era for the universe bounded by the apparent horizon as a thermodynamical system and the dark fluid is in the form of perfect fluid with constant equation of state. However, equilibrium configuartion is possible only in the phantom phase but not in the quintessence domain.

On the other hand, for universe bounded by the event horizon as a thermodynamical system, the restrictions for the validity of GSLT and maximum entropy configurations are the following:\\
$\bullet$ $\dot{S}_E+\dot{S}_{fE}\geq 0$~in quintessence era ($\rho+p>0$, {\it i.e.,} $v_A>0$) if $R_E\geq R_A$, {\it i.e.,} $v_E\geq 0$,\\
$~~~~~~~~~~~~~~~~$ $\geq 0$~in phantom era ($\rho+p<0$, {\it i.e.,} $v_A<0$) if $R_E\leq R_A$, {\it i.e.,} $v_E\leq 0$,\\
$~~~~~~~~~~~~~~~~$ $=0$~at phantom crossing ($\rho+p=0$, {\it i.e.,} $v_A=0$).\\
For equilibrium configuration $\ddot{S}_E+\ddot{S}_{fE}<0$ to be satisfied, the bounds on the radius and/or the velocity of the apparent/event horizon have been presented in Table I.\\
\begin{center} {\bf Table-I}: Constraints for equilibrium configuration for universe bounded by the event horizon \end{center}
\begin{center}
\begin{tabular}{|p{2.2cm}|p{15.4cm}|}
\hline \begin{center} Era \end{center} & \begin{center} Constraints \end{center}\\
\hline \hline \begin{center} Quintessence \end{center} & \begin{center} $R_A<R_E<\frac{2R_A}{|1+3\omega|}$, $v_A>v_E$ \end{center} \begin{center} {\it or equivalently}\end{center} \begin{center} $v_E>0$,  $v_A>\frac{v_E(1+2v_E)}{(1+v_E)^2}$\end{center}\\
\hline \begin{center} Phantom \end{center} & \begin{center} $R_E<min[R_A,\frac{2R_A}{|1+3\omega|}]$, $v_A>v_E$ or $R_E>max[R_A,\frac{2R_A}{|1+3\omega|}]$, $v_A<v_E$ and GSLT does not hold \end{center} \begin{center} {\it or equivalently} \end{center} \begin{center} $v_E<0$ and $|\frac{v_A}{v_E}|>\frac{(1+2v_E)}{(1+v_E)^2}$ or $v_E>0$ and GSLT is not satisfied \end{center}\\
\hline
\end{tabular}
\end{center}

\vspace{0.4cm}

Hence an equilibrium configuration is possible for universe bounded by the event horizon with dark perfect fluid ({\it i.e.,} $\omega<-\frac{1}{3}$) in both the quitessence and the phantom era, provided there are some bounds on the horizon radii or equivalently on the velocities of the horizons.

\begin{figure}
\includegraphics[height=4in, width=4in]{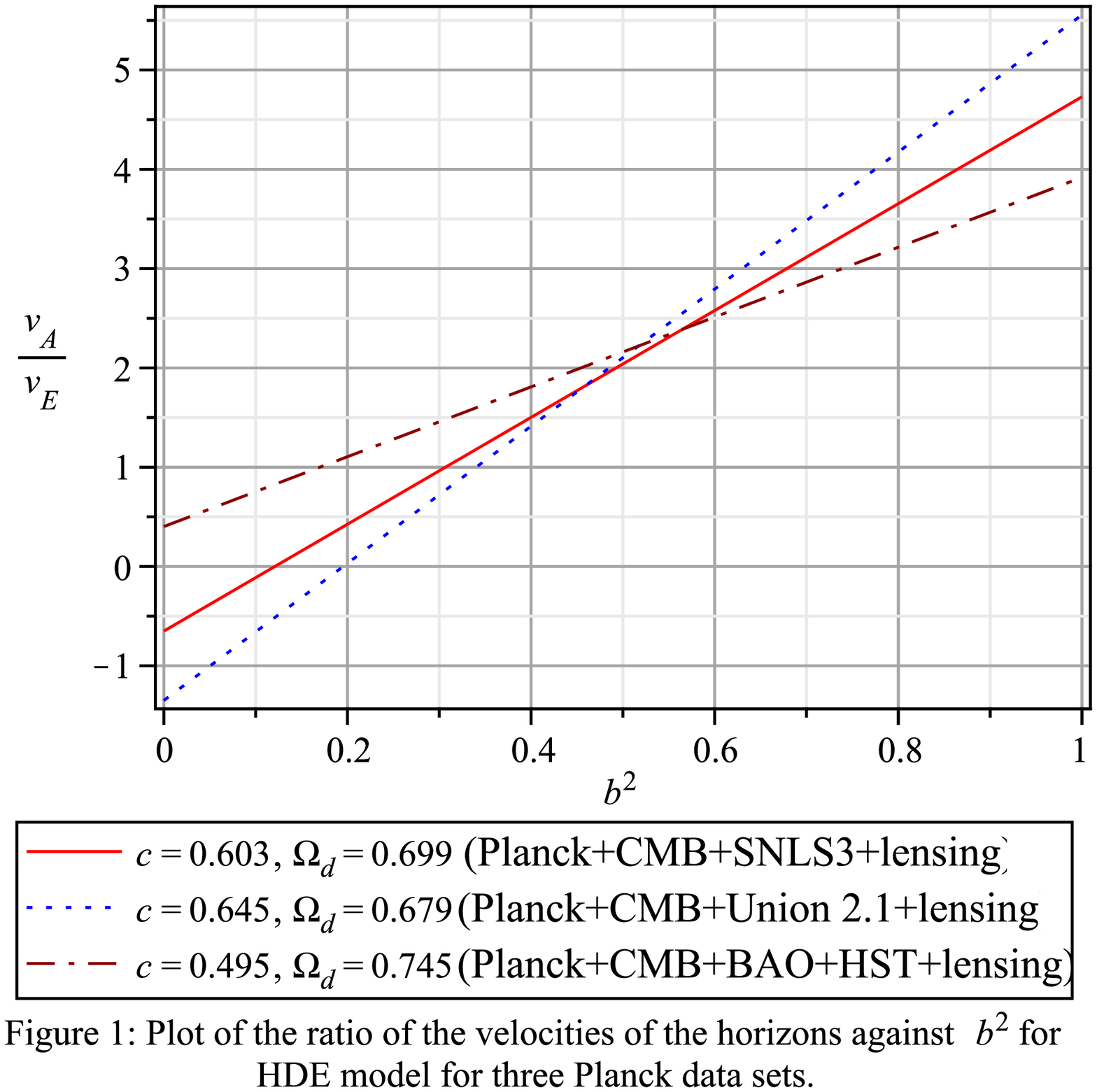}
\end{figure}

\subsection{Holographic Dark Energy (HDE) model}

In 2006, it was shown by Das {\it et al.} [13] and Amendola {\it et al.} [14] that an interaction (between holographic DE and DM which is in the form of dust) model of the universe mimicks the observationally measured phantom equation of state as compared to non-interaction models which may predict the equation of state to be of a non-phantom type. Moreover, observational data from the CMB [15] and matter distribution at large scales [16] favour the interaction models. It is therefore very natural to consider models where dark energy interacts with dark matter. 

The variable equation of state parameter for the holographic DE has the form [4,17]
\begin{equation}
\omega _d=-\frac{1}{3}-\frac{2}{3}\frac{\sqrt{\Omega _d}}{c}-\frac{b^2}{\Omega _d}
\end{equation}
where $c$ is a dimensionless parameter (estimated from the observation), the interaction term has the form $3b^2H(\rho _m+\rho _d)$ with $b^2$ as the coupling parameter between DE and DM. The density parameter $\Omega _d$ evolves as [17]

\begin{equation}
{{\Omega}^{'} _{d}}=\Omega _d [(1-\Omega _d)(1+\frac{2}{c}\sqrt{\Omega _d})-3b^2]
\end{equation}
where $'=\frac{\partial}{\partial x}$, $x=ln a$.

The velocities of the horizons can be expressed in terms of the density parameter $\Omega _d$ as
\begin{equation}
v_A=\frac{3}{2}[(1-b^2)-\frac{\Omega _d}{3}(1+\frac{2\sqrt{\Omega _d}}{c})]~~~and~~~v_E=(\frac{c}{\sqrt{\Omega _d}}-1).
\end{equation}

The ratio of the two velocities has been plotted against $b^2$ for three {\it Planck} data sets [18] in Fig. 1. The {\it Planck} results reduce the error by 30\%-60\% when compared to the {\it WMAP}-9 results. Thus, we have considered Planck+CMB data combined with the external astrophysical data sets, {\it i.e.,} the BAO measurments from 6dFGS+SDSS DR7(R)+BOSS DR9, the Hubble constant obtained directly from the {\it HST}, the supernova data set SNLS3 and Union 2.1 supernova data set as they significantly contribute to the accuracy of the constraint results. Further, we have added the lensing data as it improves the constraints by 2\%-15\%. Also, no evident tension has been found [18] when {\it Planck} data is combined with BAO, {\it HST} and Union 2.1 data sets. However, a weak tension arises when SNLS3 is combined with the other data sets. 

For thermodynamical analysis we have for the apparent horizon,
\begin{equation}
\dot{S}_A+\dot{S}_{fA}=\frac{9\pi}{2GH}(1+\omega_d \Omega_d)^2=\frac{9\pi}{2GH}[(1-b^2)-\frac{\Omega_d}{3}(1+\frac{2}{c}\sqrt{\Omega_d})]^2=\frac{2\pi}{GH} v_A^{2}
\end{equation}
and
\begin{equation}
\ddot{S}_A+\ddot{S}_{fA}=\frac{9\pi}{2G}(1+\omega_d \Omega_d)[(1+6\omega_d \Omega_d)-(5+3\omega_d \Omega_d)\frac{\dot{p}}{\dot{\rho}}]=\frac{3\pi}{G}v_A[\frac{2}{3}v_A(4v_A-5)-2(1+v_A)\frac{\dot{p}}{\dot{\rho}}],
\end{equation}
where $\frac{\dot{p}}{\dot{\rho}}=\frac{1+\frac{3}{c}\sqrt{\Omega_d}}{[(1-b^2)-\frac{\Omega_d}{3}(1+\frac{2}{c}\sqrt{\Omega_d})]}-\frac{1}{3}\lbrace{\Omega_d(1+\frac{2}{c}\sqrt{\Omega_d})+3b^2\rbrace}$.\\

As usual, the GSLT is satisfied at the apparent horizon irrespective of the fluid nature while for thermodynamical equilibrium configuration, no definite conclusion can be drawn due to complicated expression of the second derivative in Eq. (16). However, we have plotted $\ddot{S}_A+\ddot{S}_{fA}$ against the coupling parameter $b^2$ for the above three {\it Planck} data sets in Fig. 2. It is evident from the figure that equilibrium configuration for universe bounded by apparent horizon is not satisfied for any of the three {\it Planck} data sets. We have also presented bounds on $v_A$ which make $\ddot{S}_A+\ddot{S}_{fA}>0$ in Table II for some realistic choices of the coupling parameter $b^2$ considering the second equality in Eq. (16).\\

\begin{figure}
\begin{minipage}{0.4\textwidth}
\includegraphics[width=1.0\linewidth]{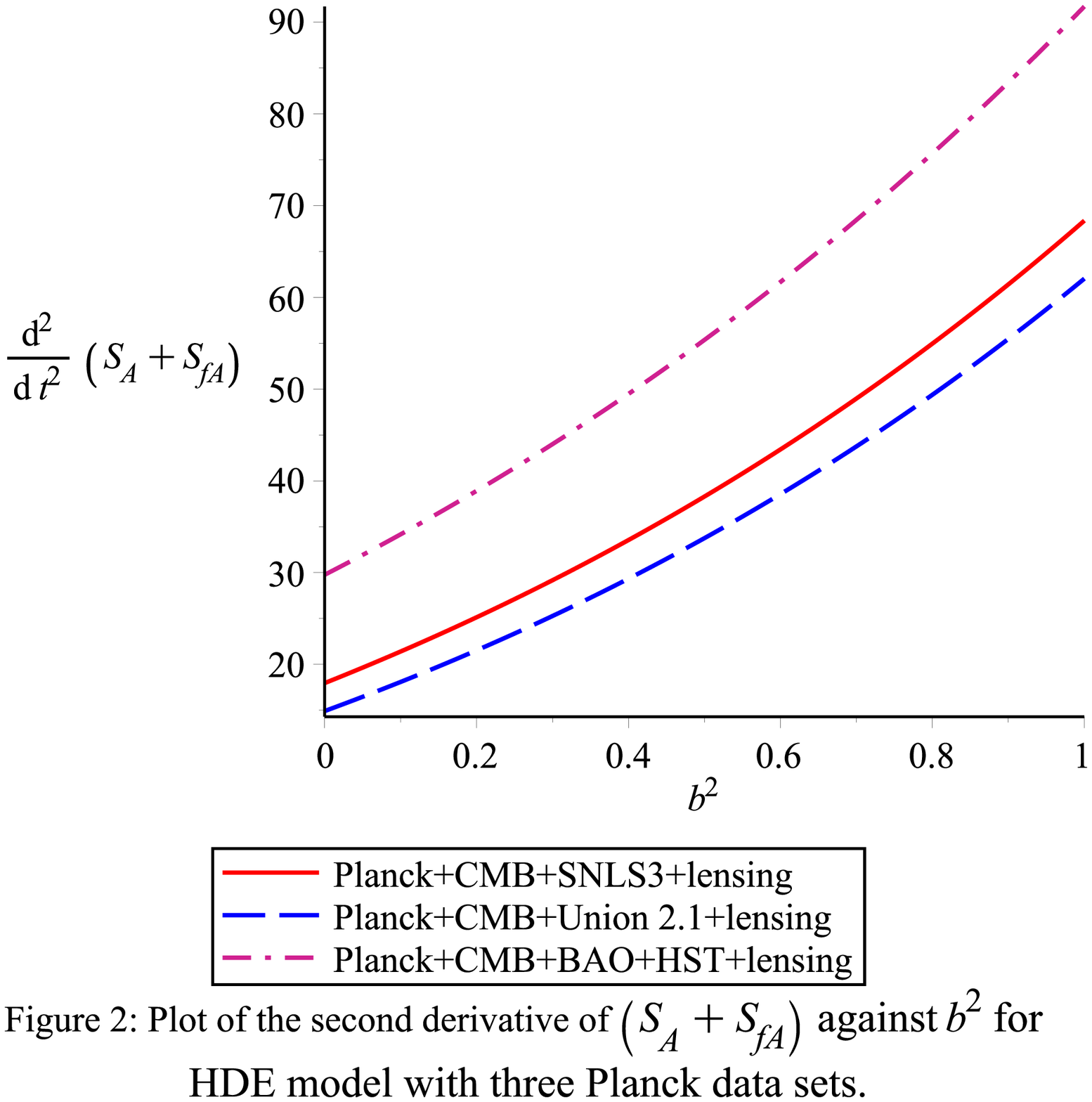}
\end{minipage}
\begin{minipage}{0.4\textwidth}
\includegraphics[width=1.0\linewidth]{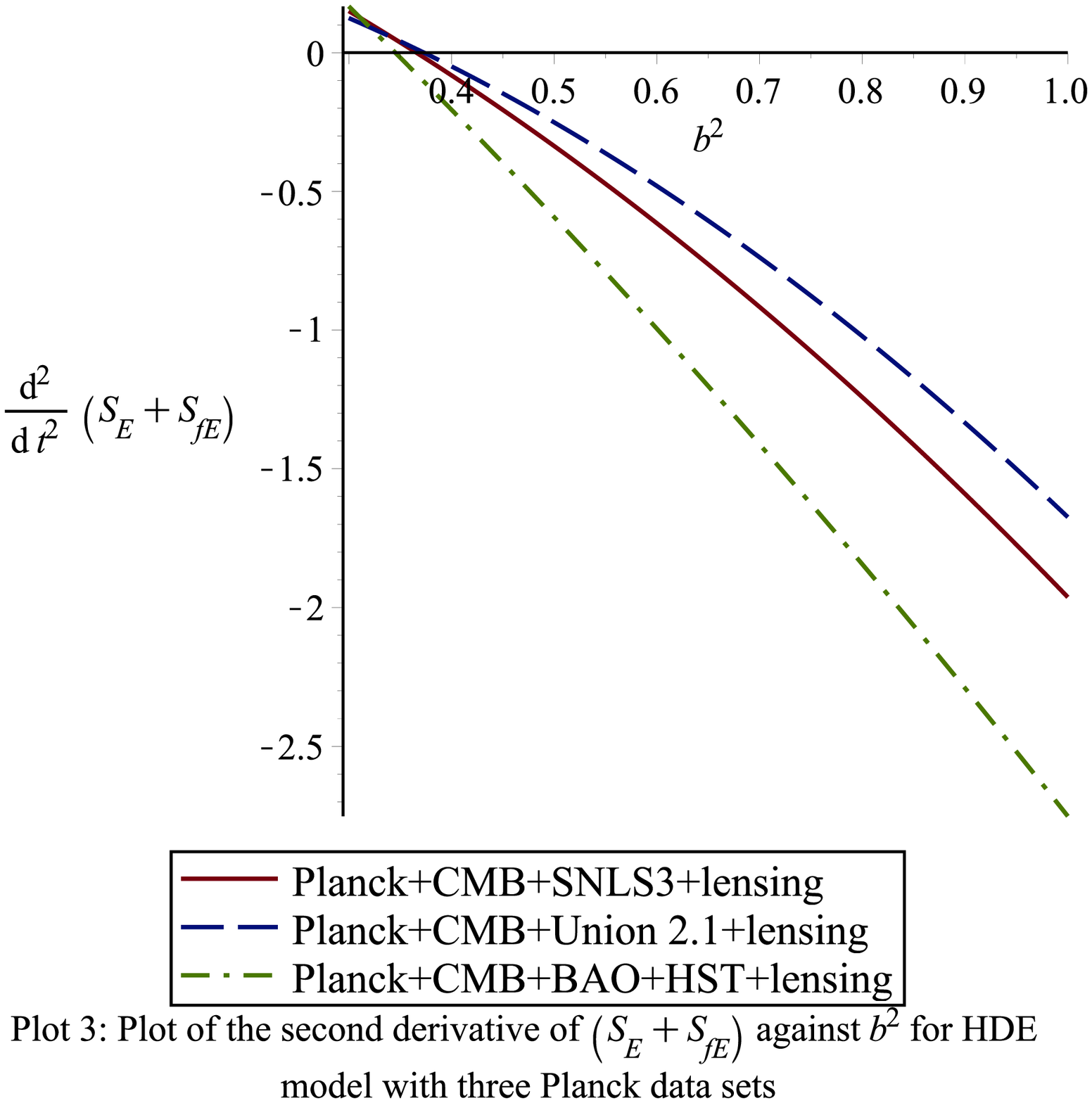}
\end{minipage}
\end{figure}

\begin{center} {\bf Table-II}: Bounds on $v_A$ for $\ddot{S}_A+\ddot{S}_{fA}>0$ for different {\it Planck} data sets and different $b^2$ \end{center} 
\begin{center}
\begin{tabular}{|c|c|c|c|c|c|}
\hline Data & $c$ & ${\Omega}_{d}$ & $b^2$ & $\frac{\dot{p}}{\dot{\rho}}$ & Bounds on $v_A$\\
\hline \hline Planck+CMB+SNLS3+lensing & 0.603 & 0.699 & 0.01 & 45.63955631 & -0.9399$<v_A<$0 {\it or} $v_A>$36.4195\\
\hline Planck+CMB+Union 2.1+lensing & 0.645 & 0.679 & 0.1 & 49.76987810 & -0.9445$<v_A<$0 {\it or} $v_A>$39.5219\\
\hline Planck+CMB+BAO+HST+lensing & 0.495 & 0.745 & 0.5 & -11.75660080 & -6.1287$<v_A<$-1.4387 {\it or} $v_A>$0\\
\hline
\end{tabular}
\end{center}

\vspace{0.4cm}

For the event horizon, we have
\begin{equation}
\dot{S}_E+\dot{S}_{fE}=\frac{2\pi R_E}{G}v_Av_E
\end{equation}
and
\begin{equation}
\ddot{S}_E+\ddot{S}_{fE}=-\frac{2\pi}{G}v_Av_E[\frac{v_A}{v_E}(1+v_E)^2-(1+2v_E)+\frac{3{\Omega}^{'} _{d}}{2v_A}(\frac{1}{3}+\frac{2}{c}\sqrt{\Omega _d})(1+v_E)],
\end{equation}
or equivalently,
\begin{equation}
\ddot{S}_E+\ddot{S}_{fE}=-\frac{2\pi}{G}v_Av_E[(\frac{v_A}{v_E}-1)(1+2v_E)+{v_A}{v_E}+\frac{3{\Omega}^{'} _{d}}{2v_A}(\frac{1}{3}+\frac{2}{c}\sqrt{\Omega _d})(1+v_E)].
\end{equation}
Now from figure 1, for $b^2 <\frac{1}{3}$, we have two possibilities ({\it approximately}):
\begin{center}
{\it either}, $0<\frac{v_A}{v_E}<1$\\
{\it or},~~$\frac{v_A}{v_E}<0$.
\end{center}
In Table III, we have considered various sign combinations of $v_A$, $v_E$ and $\Omega ^{'} _d$ and verified the validity of GSLT and thermodynamical equilibrium in each case. We have deliberately left out the case where ($v_A$,$v_E$,$\Omega ^{'}_d$) $\rightarrow$ (+ - +) as GSLT does not hold and no definite conclusion can be drawn about thermodynamical equilibrium which makes the combination unrealistic for thermodynamical study. The combinations (+ - -), (- + +) and (- + -) reveal that even though GSLT does not hold, thermodynamical equilibrium can be attained (and that too without any restriction on the model parameters). The bounds on $\Omega _d$ have also been listed against each combination. 

Again from Fig. 1, we see that $\frac{v_A}{v_E} \geq 1$ for $b^2 \geq \frac{1}{3}$ ({\it approximately}). Thus only two cases are possible here:
\begin{center}
{\it either}, $v_A >0$, $v_E >0$, ${\Omega}^{'} _{d} >0$\\
{\it or}, $v_A <0$, $v_E <0$, ${\Omega}^{'} _{d} <0$.
\end{center}
One can easily verify from Eq. (19) that in both the above cases, GSLT holds and thermodynamical equilibrium is attained. The results for this case have been presented in Table IV with bounds on $\Omega _d$. For the combinations (+ + -) and (- - +), although GSLT is satisfied, no definite conclusion is possible for equilibrium configuration and hence they are not presented in Table IV.\\

\begin{center}{\bf Table-III}: Thermodynamical conditions for $b^2<\frac{1}{3}$ for various sign combinations of ($v_A$,$v_E$,${\Omega}^{'}_{d}$)\end{center}
\begin{center}
\begin{tabular}{|c|c|c|c|c|c|}
\hline $v_A$ & $v_E$ & ${\Omega}^{'}_{d}$ & $\Omega _d$ & GSLT & Thermodynamical Equilibrium\\
\hline \hline + & + & + & $\Omega _u<\Omega_d<min(\Omega _{d_0}, c^2, \Omega_{d_1}, \Omega_v)$ & Holds & Attained\\
\hline + & + & - & $\Omega _{d_1}<\Omega _d<min(\Omega _{d_0}, c^2)$ & Holds & No definite conclusion\\
\hline + & - & - & $max(c^2, \Omega _{d_1})<\Omega _d<\Omega _{d_0}$ & Does not hold & Attained without any restriction\\
\hline - & + & + & $\Omega _{d_0}<\Omega _d<min(c^2, \Omega _{d_1})$ & Does not hold & Attained without any restriction\\
\hline - & + & - & $max(\Omega _{d_0}, \Omega _{d_1})<\Omega _d<c^2$ & Does not hold & Attained without any restriction\\
\hline - & - & + & $max(\Omega _{d_0}, c^2)<\Omega _d<\Omega_{d_1}$ & Holds & No definite conclusion\\
\hline - & - & - & $\Omega _d>max(c^2, \Omega _{d_0}, \Omega _{d_1}, \Omega_v)$ & Holds & Attained\\
\hline
\end{tabular}
\end{center}

\vspace{1cm}
\begin{center}{\bf Table-IV}: Thermodynamical conditions for $\frac{1}{3}\leq b^2<1$ for various sign combinations of ($v_A$,$v_E$,${\Omega}^{'}_{d}$)\end{center}
\begin{center}
\begin{tabular}{|p{0.4cm}|p{0.4cm}|p{0.4cm}|p{5.9cm}|p{2.5cm}|p{5.8cm}|}
\hline  ${v_A}$ & ${v_E}$ & ${\Omega}^{'}_{d}$ & \hspace{2.7cm}$\Omega _d$ & ~~~~~~GSLT & ~~Thermodynamical Equilibrium\\
\hline \hline  + & + & + & \hspace{0.77cm}$\Omega_\alpha$$<$$\Omega _d$$<$$min(\Omega_{d_0}, c^2, \Omega_\beta)$ & ~~~~~~Holds & ~~~~~~~~~~~~~~~~Attained \\
\hline  ~- & ~- & ~-& \begin{center}$either$ $\Omega _d$$>$$max(\Omega _{d_0}, c^2, \Omega _\beta)$\end{center} \begin{center}$or$\end{center} \begin{center}$max(\Omega_{d_0}, c^2)<\Omega_d<\Omega_\alpha$\end{center} & ~~~~~~Holds & ~~~~~~~~~~~~~~~~Attained \\
\hline
\end{tabular}
\end{center}
\vspace{0.5cm}

In the above tables $\Omega_{d_0}={x_0}^2$, where $x_0$ is the positive root of the cubic equation $2x^{3}+cx^{2}-3c(1-b^{2})=0$. Similarly, for $0<b^{2}<\frac{1}{3}$, $\Omega_{d_1}={x_1}^2$, where $x_1$ is the positive root of the cubic equation $2x^3+cx^2-2x-c(1-3b^2)=0$, while for $\frac{1}{3}<b^2<1$, this cubic equation has two positive real roots $\alpha$ and $\beta$ $(\alpha<\beta)$ and $\Omega_\alpha=\alpha ^2$ and $\Omega_\beta=\beta ^2$. In Table III, $\Omega_u=u^2, \Omega_v=v^2$ and ($u,v$) are the positive roots of the cubic equation $cx^3+(1+\frac{c^2}{2})x^2-3cx+\frac{c^2}{2}(1+3b^2)=0$. Thus we see from the above Tables III and IV that, with some restrictions on the density parameter for holographic dark energy (HDE), it is possible to satisfy both the GSLT and the thermodynamical equilibrium for universe bounded by the event horizon. On the other hand, to examine the thermodynamical equilibrium configuration for universe bounded by the event horizon from the observational point of view, we have plotted the function $\ddot{S}_E+\ddot{S}_{fE}$ against $b^2$ for the values of $\Omega _d$ and $c$ obtained from the above {\it Planck} data sets in fig. 3. The graphical representation shows that equilibrium configuration is possible for $b^2 >\frac{1}{3}$ ({\it approx.}) for all the three {\it Planck} data sets.

\subsection{Modified Chaplygin Gas}

The transition of our universe from a decelerated phase to the present epoch of cosmic acceleration is smoothly described by the Chaplygin Gas models. These models also attempt to provide a unified macroscopic phenomenological description of dark energy and dark matter as well as represent, perhaps, the simplest modification of $\Lambda$CDM models. However, for the present thermodynamical analysis, it might behave as a DE fluid.

The equation of state for modified chaplygin gas (MCG) is given by [19]
\begin{equation}
p=\gamma\rho-\frac{B}{\rho ^n},
\end{equation}
where $\gamma\leq 1$, $B>0$, $n>0$ are the parameters of the model. If we choose $\gamma=\frac{1}{3}$, this model describes the radiation era $p=\frac{1}{3}\rho$ at the very early stages of the evolution and gradually with the evolution of the universe it enters into matter dominated era and then pressure becomes negative and finally it matches to the $\Lambda$CDM model $p=-\rho$ i.e., the MCG model is extended upto phantom barrier and phantom region is forbidden for this model. So if universe filled with MCG and bounded by apparent/event horizon is an isolated thermodynamical system, then the radius and velocity of the apparent horizon are given by:
\begin{equation}
R_A=R_0\frac{a^\frac{3}{2}(1+\gamma)}{[Ba^\mu +C]^\frac{1}{2(n+1)}}~~~,~~~v_A=\frac{3}{2}\frac{C(1+\gamma)}{(Ba^\mu +C)},
\end{equation}
where $a$ is the scale factor of the FRW model, $C$ is a constant of integration, $\mu=3(1+n)(1+\gamma)$ and $R_0=\sqrt{\frac{3}{8\pi G}}(1+\gamma)^\frac{1}{2(n+1)}$. On the other hand, the radius of the event horizon can be expressed in terms of Hypergeometric function as
\begin{equation}
R_E=R_1~_2F_1[\frac{1}{2(n+1)}, \frac{1}{\mu}, 1+\frac{1}{\mu}, \frac{-C}{Ba^\mu}],
\end{equation}
with $R_1=\frac{R_0}{B^\frac{1}{2(n+1)}}$. Further, the sound speed $(c_s)$ is related to the velocity of the apparent horizon by the relation
\begin{equation}
c^{2}_s=\frac{\partial p}{\partial \rho}=\gamma _0-\frac{2}{3}nv_A,
\end{equation}

\begin{figure}
\includegraphics[height=4in, width=4in]{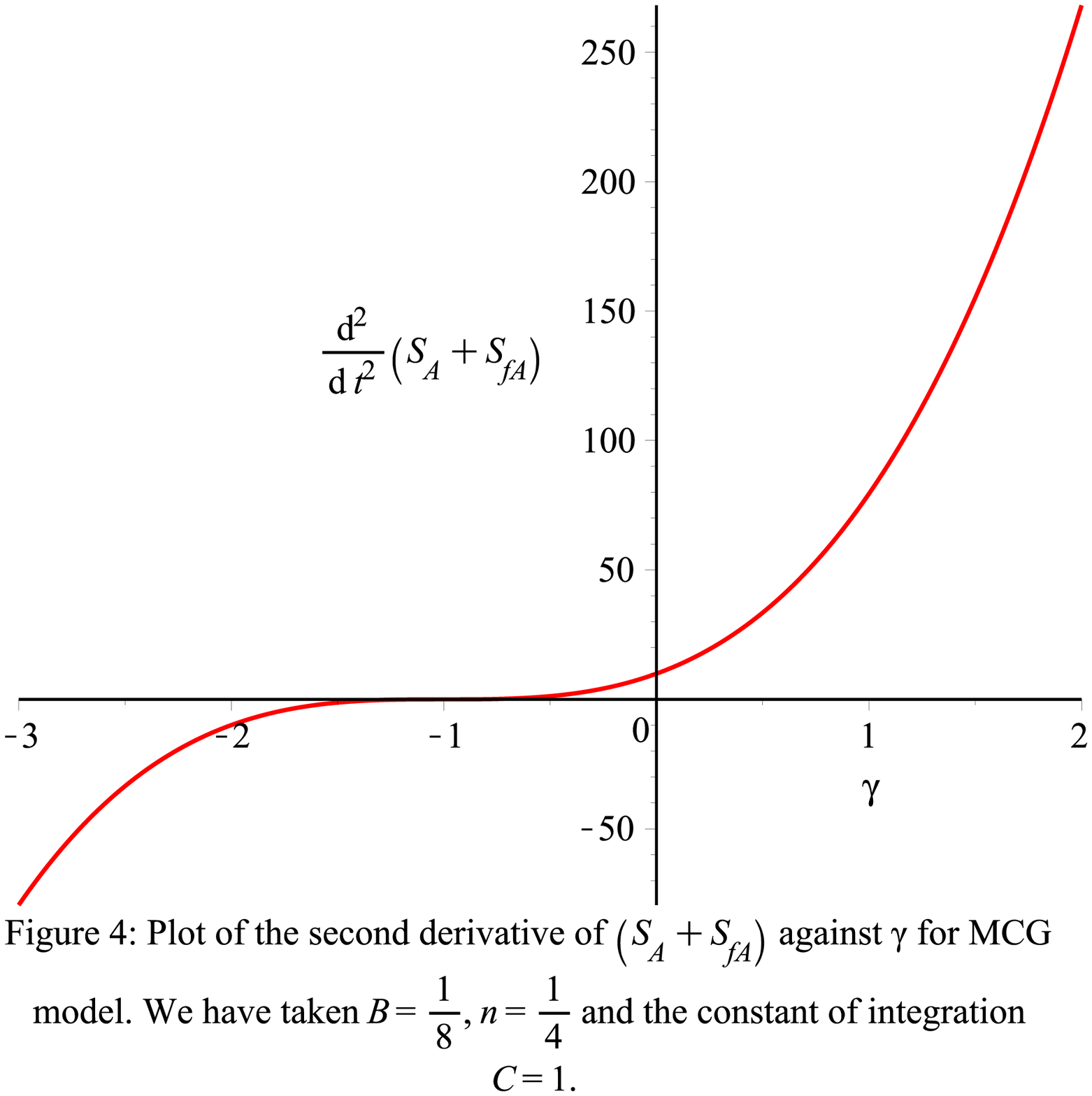}
\end{figure}

\begin{figure}
\begin{minipage}{0.4\textwidth}
\includegraphics[width=1.0\linewidth]{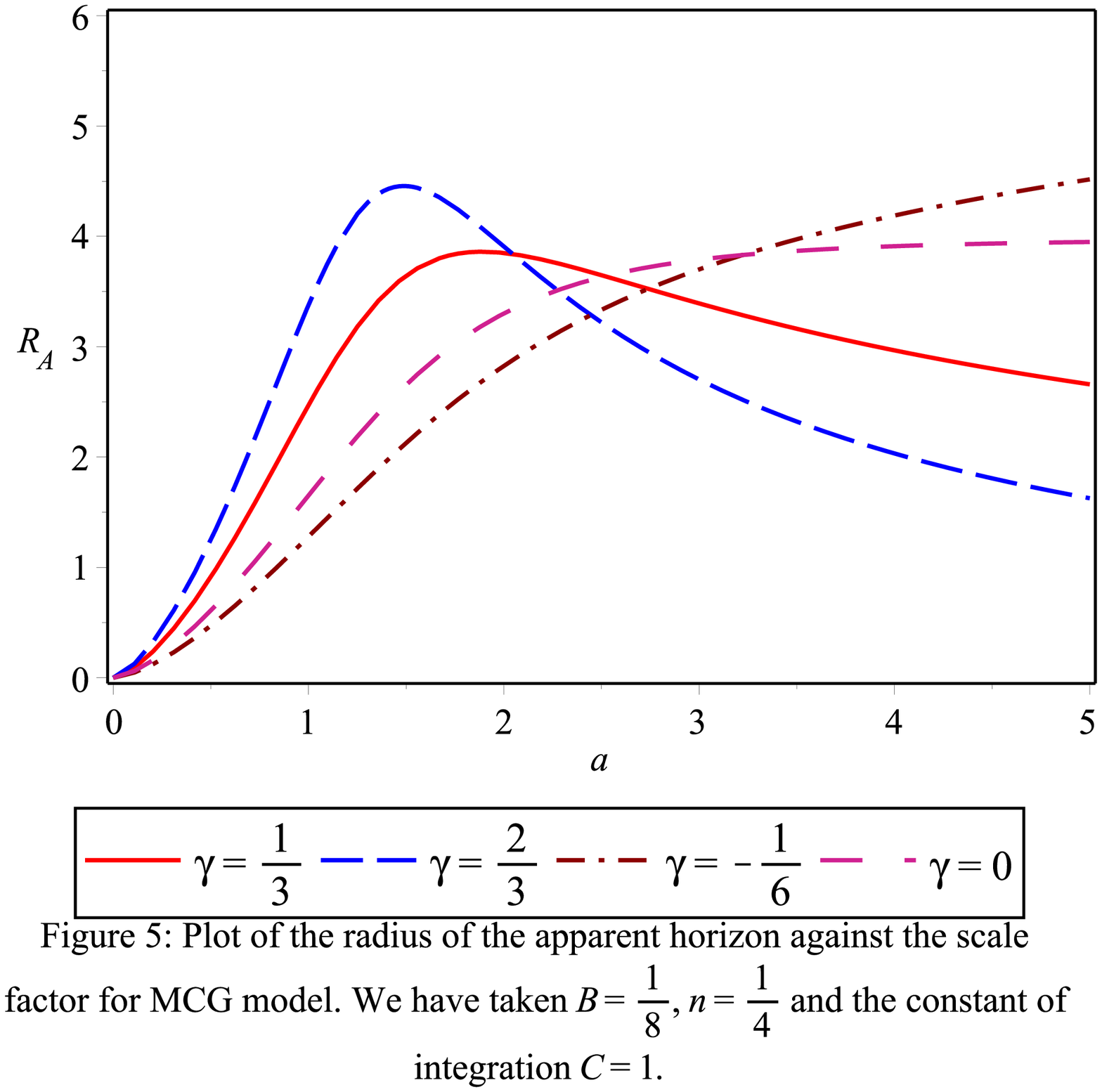}
\end{minipage}
\begin{minipage}{0.4\textwidth}
\includegraphics[width=1.0\linewidth]{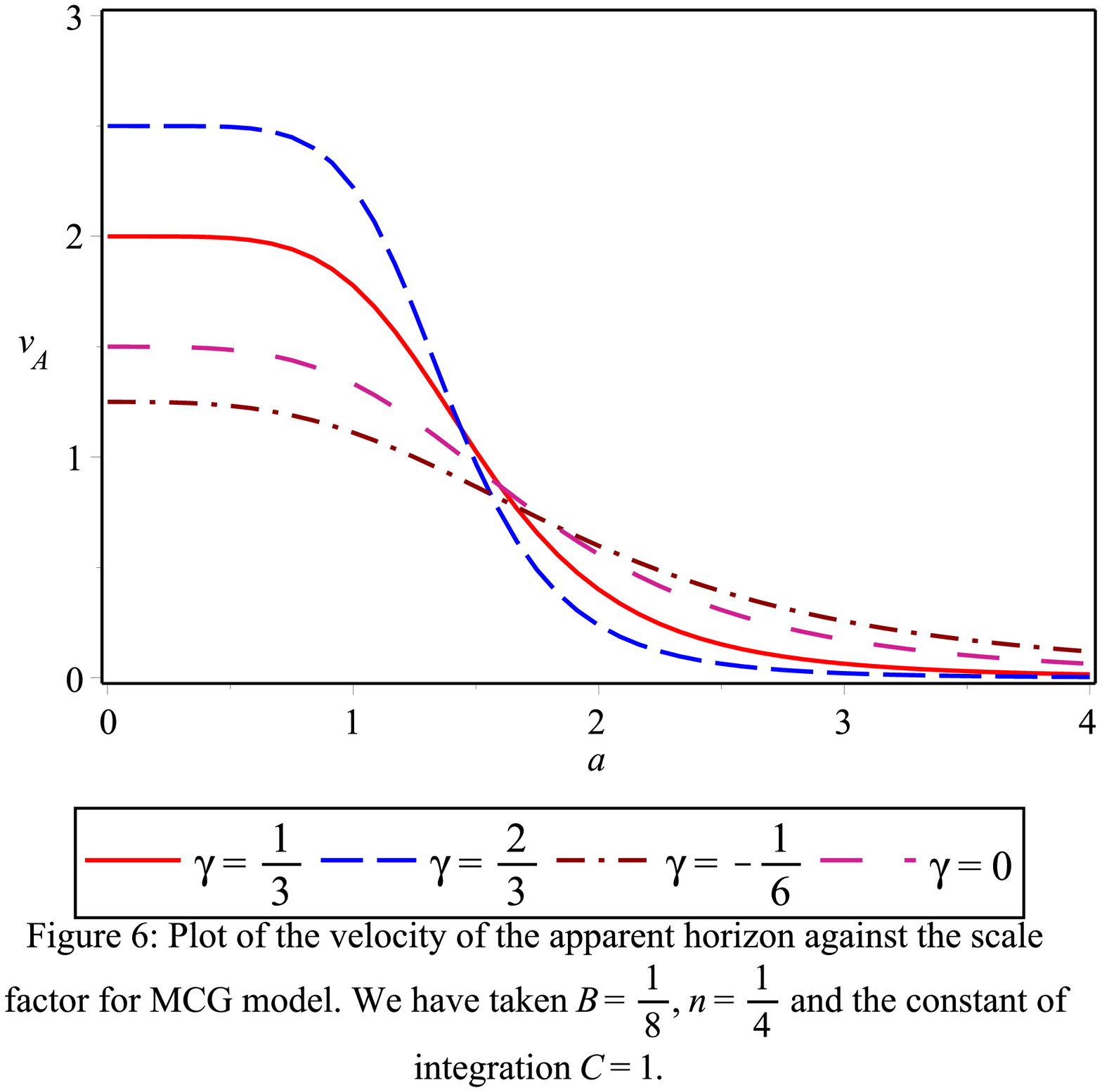}
\end{minipage}
\end{figure}
where $\gamma _0=\gamma +n(1+\gamma)>0$. Thus for the total entropy variation and thermodynamical equilibrium configuration, we have for the apparent horizon,
\begin{equation}
\dot{S}_A+\dot{S}_{fA}=\frac{9\pi}{2GH}(1+\frac{p}{\rho})^2=\frac{9\pi}{2GH}\frac{C^2(1+\gamma)^2}{(C+Ba^\mu)^2}=\frac{2\pi}{GH}{v_A}^2
\end{equation}
and
\begin{equation}
\fontsize{5pt}{6.0}
\ddot{S}_A+\ddot{S}_{fA}=\frac{9\pi}{2G}(1+\frac{p}{\rho})[(1+6\frac{p}{\rho})(1+\frac{p}{\rho})-(5+3\frac{p}{\rho}) c_s^{2}]=\frac{9\pi}{2G} \lbrace{\frac{C(1+\gamma)}{C+Ba^\mu}\rbrace}^3[C^2(1+2\gamma +3 {\gamma}^2)-BC a^\mu \lbrace{(5+7\gamma)+\gamma _0(5+3\gamma)-2\gamma _0 B^2 a^{2\mu}\rbrace}]=\frac{6\pi}{G}v_A[(\frac{4}{3} v_A^{2}-c_s^{2})-2v_A(\frac{5}{3}+c_s^{2})].
\end{equation}

\begin{figure}
\begin{minipage}{0.4\textwidth}
\includegraphics[width=1.0\linewidth]{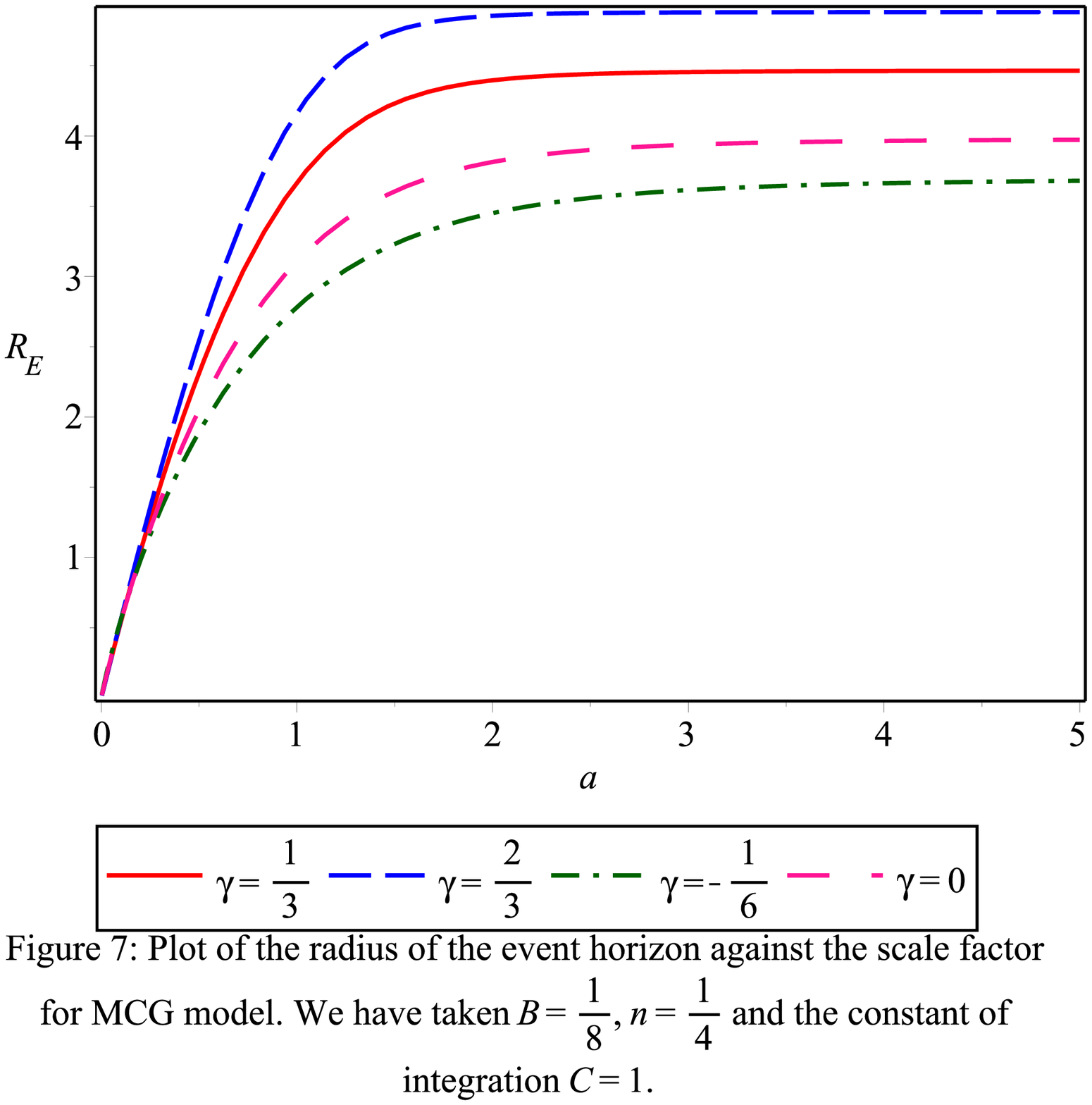}
\end{minipage}
\begin{minipage}{0.4\textwidth}
\includegraphics[width=1.0\linewidth]{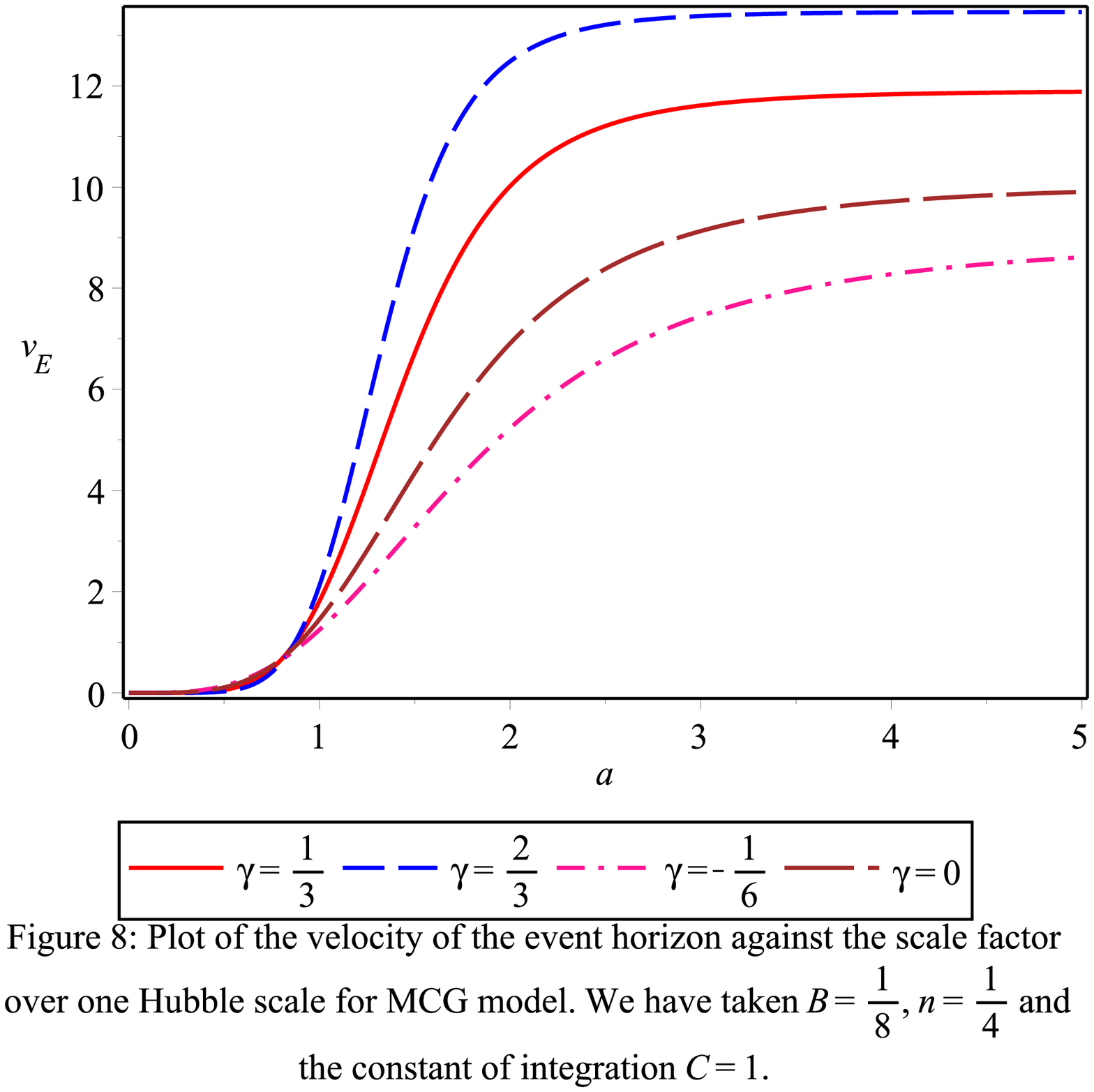}
\end{minipage}
\end{figure}
As before, GSLT holds throughout the evolution of the universe bounded by the apparent horizon while due to complicated expression, no definite conclusion can be made regarding equilibrium thermodynamical configuration. However, as before, Fig. 4 shows the graphical representation of $\ddot{S}_A+\ddot{S}_{fA}$ for MCG model against the parameter $\gamma$ and it is evident from the figure that equilibrium configuration is violated for $\gamma \geq -1$. We have also calculated the bounds on $v_A$ which make $\ddot{S}_A+\ddot{S}_{fA}>$0 and they are presented in Table V. We have considered four different values of $\gamma$ to evaluate the bounds with $B=\frac{1}{8}$, $n=\frac{1}{4}$ and the integration constant $C=1$. The radius and the velocity of the apparent horizon have been plotted against the scale factor for these values of the parameters in Fig. 5 and Fig. 6 respectively. Fig. 6 shows that the bounds on $v_A$ in Table V will hold at early and late epoch of the evolution of the universe. 

\begin{center} {\bf Table-V}: Bounds on $v_A$ for $\ddot{S}_A+\ddot{S}_{fA}>0$ for different $\gamma$ and ($B$,$n$,$C$)=($\frac{1}{8}$,$\frac{1}{4}$,1) \end{center} 
\begin{center}
\begin{tabular}{|c|c|}
\hline $\gamma$ & Bounds on $v_A$\\
\hline \hline $\frac{1}{3}$ &  $-0.0839<v_A<0$ {\it or} $v_A>2.3839$\\
\hline $\frac{2}{3}$ & $-0.1408<v_A<0$ {\it or} $v_A>2.8408$\\
\hline -$\frac{1}{6}$ & $0<v_A<0.0610$ {\it or} $v_A>1.6390$\\
\hline 0 & $v_A>1.9$\\
\hline
\end{tabular}
\end{center}

For the event horizon, we have
\begin{equation}
\dot{S}_E+\dot{S}_{fE}=\frac{8\pi ^2 R_E(\rho +p)(HR_E-1)}{H^2}=\frac{2\pi R_E}{G}v_A v_E
\end{equation}
and 
\begin{equation}
\fontsize{7pt}{8.4}
\ddot{S}_E+\ddot{S}_{fE}=8\pi ^2 (\rho +p)(R_E-\frac{1}{H})[-\lbrace{\frac{R_E}{2}(1-\frac{3p}{\rho})+\frac{1}{H}+3R_E c_s^{2}\rbrace}+R_E\lbrace{1-\frac{3(1+\frac{p}{\rho})}{2(HR_E-1)}\rbrace}]=-\frac{2\pi}{G}v_A v_E[HR_E\lbrace{(2+3\gamma _0)-(1+2n)v_A+\frac{v_A}{v_E}\rbrace}-v_E].
\end{equation}

\begin{figure}
\includegraphics[height=4in, width=4in]{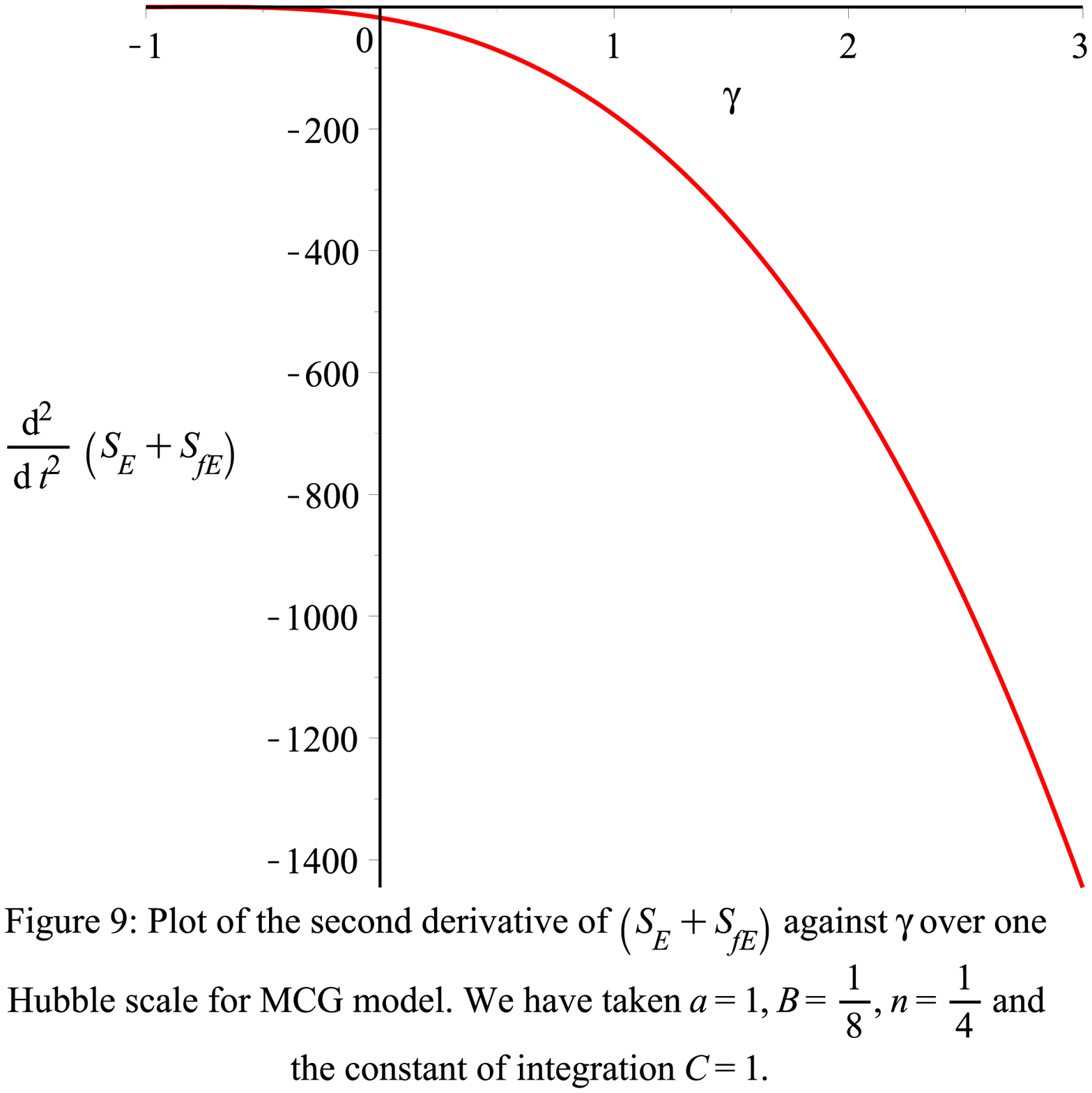}
\end{figure}

So for the validity of GSLT for the universe bounded by the event horizon, both the horizon velocities must have the same sign. In other words, in quintessence era, GSLT holds if $v_E>0$ ({\it i.e.,} $R_E>R_A$). However, for equilibrium configuration ({\it i.e., $\ddot{S}_E+\ddot{S}_{fE} <0$}), we have an upper bound on $v_E$ namely $v_E<\frac{1}{1+2n}$, {\it i.e.,} $R_E<\frac{2(1+n)}{1+2n} R_A$. Also form Eq. (22), to have a realistic sound speed ($0<{c}^2_s<1$), $v_A$ and $\gamma$ are restricted as $v_A\in(\frac{3(\gamma _0-1)}{2n}, \frac{3\gamma _0}{2n})$ and $\gamma\in(\frac{-n}{n+1}, \frac{-n}{3(n+1)})$. Now as $n>0$ so $\gamma$ should be $>-\frac{1}{3}$, {\it i.e.,} MCG might be of DE nature. Further, if $\gamma _0<1$ then the restriction on $v_A$ holds for all values of $a$, {\it i.e.,} it holds throughout the evolution of the universe while if $\gamma _0>1$ then the restriction on $v_A$ implies the scale factor to be bounded by $[\frac{C(1-\gamma)}{B(\gamma _0-1)}]^{\frac{1}{\mu}}$. In Figs. 7 and 8, the evolution of the radius of the event horizon and the velocity of the event horizon respectively, has been presented and it is clear from the figures that both $R_E$ and $v_E$ increase sharply at the early phases of the evolution of the universe and then they approach a constant value at late times. Further, Figs. 6 and 8 show that $v_A$ and $v_E$ are positive throughout the evolution, so GSLT is always satisfied in the case of event horizon. We have also shown $\ddot{S}_E+\ddot{S}_{fE}$ graphically for varying $\gamma$ in Fig. 9. From the figure, it can be concluded that thermodynamical equilibrium configuration is possible for universe bounded by the event horizon. It should be noted that in the early epochs, the MCG model behaves as a perfect fluid having equation of state $p=\gamma \rho$, so the choices $\gamma =\frac{2}{3}$ and $\gamma =\frac{1}{3}$ correspond to ultra relativistic and radiation era respectively while $\gamma =0$ stands for generalized Chaplygin gas model and $\gamma =-\frac{1}{6}$ indicates dark energy state. The choices of the other parameters are arbitrary, consistent with the restrictions $B>0$ and $n>0$ (mentioned above). Finally, it must be mentioned here that in deriving Eqs. (9), (15) and (20), we have assumed that the fluid inside the horizon has the same temperature as that of the horizon itself.

\section{CONCLUSIONS}

This paper deals with an extensive study of equilibrium thermodynamics for three different choices of DE models namely (a) Perfect fluid with constant equation of state, (b) Holographic dark energy (HDE) model and (c) Modified Chaplygin gas (MCG) model. The generalized second law of thermodynamics (GSLT) is always satisfied for universe bounded by the apparent horizon while for the event horizon, GSLT is satisfied with some restrictions on the radii/velocities of the horizons or on the fluid parameters. However, from the point of view of thermodynamical equilibrium configuration, apparent horizon seems to be less favourable than event horizon at least for these three choices of DE models. For perfect fluid with constant equation of state, equilibrium configuration holds only in the phantom domain for apparent horizon while it is possible to achieve equilibrium configuration both in the quintessence era and in the phantom era for event horizon with some realistic restrictions ({\it see} Table I). In the case of HDE model, the condition(s) for equilibrium configuration to hold good is/are very complicated for both the horizons ({\it see} Eqs. (16) {\it and} (19)). So, based on three different choices of {\it Planck} data sets, we have plotted the second order derivative of the total entropy for both the horizons ({\it see} Figs. 2 {\it and} 3) and it is found that equilibrium condition is violated for apparent horizon for all choices of the coupling parameter $b^2$ while in the case of event horizon, equilibrium configuration is attained for $b^2 >\frac{1}{3}$ ({\it approx.}). Also, in the case of event horizon, we have presented in Tables III and IV, the bounds on the density parameter for the validity of thermodynamical equilibrium. In the case of MCG model, graphical representation again shows that equilibrium configuration is not possible for apparent horizon with $\gamma \geq -1$ while it is possible to have equilibrium configuration in the case of event horizon for $\gamma \geq -1$. Thus universe bounded by the event horizon is a perfect thermodynamical system (second law of thermodynamics is valid for the system and thermodynamical equilibrium is attained) for the above DE models and hence it is reasonable to assume that the temperature of the fluid is same as that of the horizon. Therefore, we may conclude that DE (which is predicted by recent observations) favurs event horizon for universal thermodynamics.


\begin{acknowledgments}

One of the authors (S.C.) is thankful to IUCAA, Pune, India for their warm hospitality as a part of the work has been done during a visit. Also S.C. is thankful to DRS programme in the Department of Mathematics, Jadavpur University. The author S.S. is thankful to UGC-BSR Programme of Jadavpur University for awarding JRF. The authors are also thankful to the anonymous referee(s) for their constructive comments and illuminating suggestions.

\end{acknowledgments}


\frenchspacing
\begin{thebibliography}{19}

\bibitem{Riess1} A.G. Riess {\it et al}, {\it Astron. J.} {\bf 116}, 1009(1998); S. Perlmutter et al. {\it Astrophys. J.} {\bf 517} 565(1999).
\bibitem{Spergel1} D.N. Spergel {\it et al}, {\it Astrophys. J. Suppl. Ser.} {\bf 148} 175(2003); {\bf 170} 377(2007).
\bibitem{Tegmark1} M. Tegmark {\it et al}, {\it Phys. Rev. D} {\bf 69} 103501(2004); D.J. Eisenstein et al. {\it Astrophys. J.} {\bf 633} 560(2005).
\bibitem{Wang1} B. Wang, Y. Gong and E. Abdalla, {\it Phys. Rev. D}  {\bf 74} 083520(2006).
\bibitem{Mazumder1} N. Mazumder and S. Chakraborty, {\it  Class. Quant. Grav.}  {\bf 26} 195016(2009); {\it  Gen. Rel. Grav.}  {\bf 42} 813(2010);   {\it  Eur. Phys. J. C}  {\bf 70} 329(2010);   S. Chakraborty, N. Mazumder and R. Biswas {\it Eur. Phys. Lett.}  {\bf 91} 4007(2010); {\it  Gen. Rel. Grav.}  {\bf 43} 1827(2011); J. Dutta and S. Chakraborty  {\it  Gen. Rel. Grav.}  {\bf 42} 1863(2010).
\bibitem{Chakraborty} S. Chakraborty,   {\it Phys. Lett. B} {\bf 718} 276(2012). 
\bibitem{Pavon1} D. Pavon and W. Zimdahl,  {\it Phys. Lett. B} {\bf 708} 217(2012).
\bibitem{Saha1} S. Saha and S. Chakraborty, {\it Phys. Lett. B} {\bf 717} 319(2012).
\bibitem{Radicella1} N. Radicella and D. Pavon, {\it Gen. Rel. Grav} {\bf 44} 685(2012).
\bibitem{Callen} H.B. Callen, {\it Thermodynamics} (J. Wiley, N.Y., 1960).
\bibitem{Izquierdo1} G. Izquierdo and D. Pavon,   {\it Phys. Lett. B}  {\bf 633} 420(2006).
\bibitem{Bennet1} C.L. Bennet {\it et al}, {\it Astron. and Astrophy.}  {\bf 399} L19(2003);  {\bf 399} L25(2003);   P. de Bernardis {\it et al},  {\it Nature}  {\bf 404} 955(2000); Gold {\it et al},  {\it APJs}  {\bf 192} 15(2011).
\bibitem{Das1} S. Das, P.S. Corasaniti and J. Khoury, {\it Phys. Rev. D} {\bf 73} 083509(2006).
\bibitem{Amendola1} L. Amendola, M. Gasperini and F. Plazza, {\it Phys. Rev. D} {\bf 74} 127302(2006).
\bibitem{Olivares1} G. Olivares, F. Atrio-Barandela and D. Pavon, {\it Phys. Rev. D} {\bf 71} 063523(2005).
\bibitem{Olivares1} G. Olivares, F. Atrio-Barandela and D. Pavon, {\it Phys. Rev. D} {\bf 74} 043521(2006).
\bibitem{Wang1} B. Wang, Y. Gong and E. Abdalla, {\it Phys. Rev. D}  {\bf 624} 141(2005).
\bibitem{Li1} M. Li, X.D. Li, Y.J. Ma, X. Zhang and Z. Zhang {\it arxiv:} 1305.5302/astro-ph.CO.
\bibitem{Debnath1} U. Debnath, A. Banerjee, S. Chakraborty, {\it Class. Quant. Grav.} {\bf 21} 5609(2004).
\end{thebibliography}
\end{document}